\RequirePackage{lineno} 
\documentclass{emulateapj}
\usepackage{times}
\usepackage{graphicx}

\shorttitle{SMBHB 4C+37.11}
\shortauthors{Romani et al.}

\begin{document}

\title{A Multi-wavelength Study of the Host Environment of SMBHB 4C+37.11}

\author{Roger W. Romani\altaffilmark{1}, W.~R. Forman\altaffilmark{2}, Christine Jones\altaffilmark{2},
S.~S. Murray\altaffilmark{3}, A.~C. Readhead\altaffilmark{4}, 
Greg B. Taylor\altaffilmark{5},
and R.~T. Zavala\altaffilmark{6,}\altaffilmark{7}}
\altaffiltext{1}{Department of Physics, Stanford University, Stanford, CA 94305, USA \email{rwr@astro.stanford.edu}}
\altaffiltext{2}{Smithsonian Astrophysical Observatory, 60 Garden Street, Cambridge, MA 02138, USA}
\altaffiltext{3}{Johns Hopkins University, Baltimore, MD, USA}
\altaffiltext{4}{Cahill Center for Astronomy and Astrophysics, California Institute of Technology 
1200 E. California Blvd, Pasadena, CA 91125, USA}
\altaffiltext{5}{Department of Physics and Astronomy, University of New Mexico, MSC07 4220, 
Albuquerque, NM 87131-0001, USA}
\altaffiltext{6}{U.S. Naval Observatory, Flagstaff Station, 10391 W. Naval Obs. Rd., Flagstaff, AZ 86001}
\altaffiltext{7}{Visiting Astronomer, Kitt Peak National Observatory, which is operated by the Association 
of Universities for Research in Astronomy (AURA) under cooperative agreement with the National Science Foundation.
The WIYN Observatory is a joint facility of the University of Wisconsin-Madison, Indiana University, Yale University, and the National Optical Astronomy Observatory.}

\begin{abstract}

	4C+37.11, at z=0.055 shows two compact radio nuclei, imaged by VLBI at 7mas
separation, making it the closest known resolved super-massive black hole binary (SMBHB).
An important question is whether this unique object is young,
caught on the way to a gravitational in-spiral and merger, or has `stalled'
at 7pc. We describe new radio/optical/X-ray observations of the massive host and its
surrounding X-ray halo. These data reveal X-ray/optical channels following the
radio outflow and large scale edges in the X-ray halo. These structures are promising
targets for further study which should elucidate their relationship to the unique SMBHB core.
\end{abstract}


\section{Introduction}

	Super-massive black hole binaries (SMBHB) are expected to form during hierarchical
assembly of large galaxies \citep{bbr}. Yet, while the evidence for major mergers is now
strong (e.g. Bell et al. 2006), the evidence for SMBHB is much more limited. Some double
X-ray nuclei have been identified at kpc-scale separations \citep[e.g.][]{owen85,kom03}, but
at these scales the holes may not be bound by mutual gravitational interaction. Indirect
evidence for binarity at small separation has also been inferred from QSO emission line structure
\citep[e.g.][]{BL09}, although other interpretations remain viable in many cases
\citep[][and references therein]{era12}. To date, the only spatially resolved SMBHB with 
a parsec scale separation is the nucleus of the Compact Symmetric Object (CSO) radio galaxy 4C+37.11. 
This galaxy contains a pair of unresolved, variable, non-thermal radio sources with a 
projected separation of 7 pc \citep{man04,rod06}.

	Attempts to find other such systems have not yet yielded fruit. For example,
\citet{BS11}, in a review of multi-frequency VLBI maps of over 3000 AGN, confirmed
that only 4C+37.11 shows strong evidence for a resolved double core. Evidently, the rarity
of such objects means that transition is relatively rapid from kpc-scale binaries to sub-pc 
scale where the merger lifetime to gravitational radiation loses is guaranteed to be short. Given the
inadequacy of simple dynamical friction \citep{bbr}, this raises some interesting
theoretical questions. It also has important observational implications
for the energy injection into the host nucleus \citep{mm03,spr05} and for the background of
ultra-low frequency gravitational radiation that may be detectable with space-based
\citep{hug03} or pulsar timing \citep{det79} searches. More specifically, it leads us to ask
why 4C+37.11 is uniquely seen at small separation. If not simple observational selection or
fortuitous timing,
perhaps some peculiarity of the system has delayed merger. Given the strong expected feedback
between the AGN nucleus and the host environment \citep{cet09}, we can hope that study
of 4C+37.11 environment may yield useful clues.

	In our initial study of the system we were only able to supplement the cm-wavelength
VLBI maps with limited multi-wavelength data. For the optical we had only sky survey imaging,
a shallow R-band frame in \citet{skf93}, and spatially unresolved Hobby-Eberly Telescope (HET) 
spectroscopy. However, 4C+37.11 was detected as a resolved X-ray source by {\it{ROSAT}} \citep{rosat},
suggesting an unique (for a CSO) host environment.  The situation called for a deeper multi-wavelength study.
Here we report on observations of the host spanning the radio to x-ray regimes.
These observations (\S2) were obtained with the Karl G. Jansky Very Large Array (VLA), the WIYN 3.5\,m 
telescope (optical imaging), the W. M. Keck I 10\,m telescope (optical 
spectra) and the {\it{Chandra}} X-ray Observatory (x-ray imaging spectroscopy). Our analysis has indeed uncovered an 
unusual host morphology and evidence for AGN-host feedback. However further work will be needed to
obtain definitive constraints on the merger age and current evolution of the SMBHB.

\begin{figure*}[ht!!]
\vskip 7.9truecm
\includegraphics{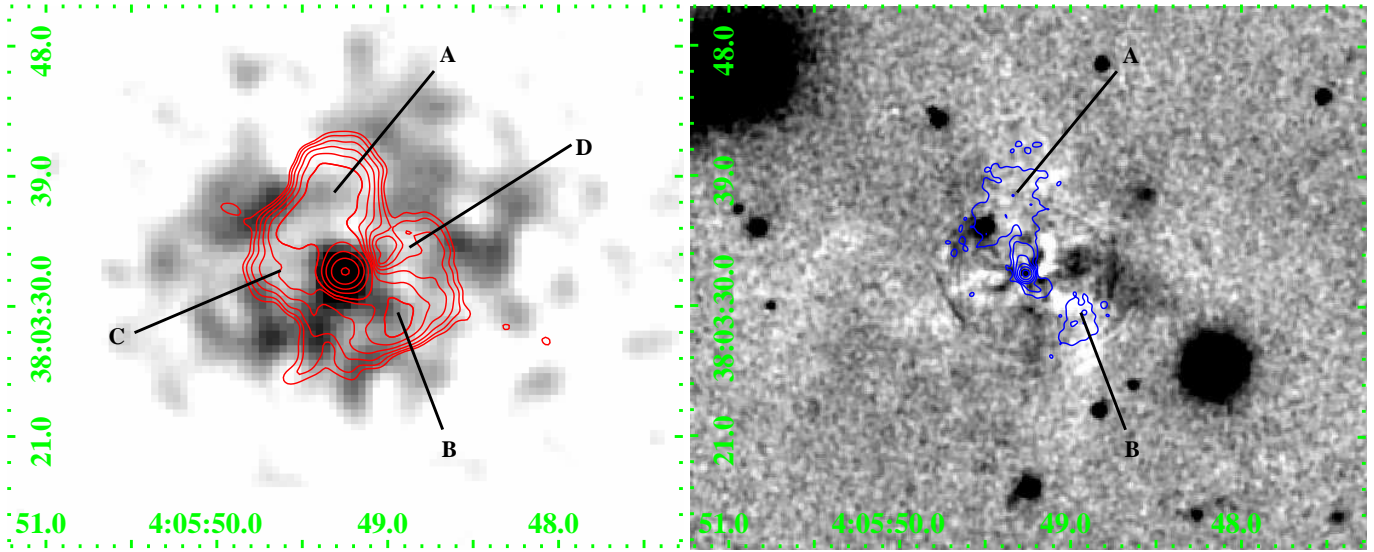}
\begin{center}
\caption{\label{Radio} 
Left: Contours of the VLA 1.4\,GHz map, showing the currently active jet
axis (A,B). An apparent older bipolar outflow at nearly orthogonal position angle 
is also present (C,D). The background gray-scale is the central region of
the CXO 0.3-8\,kev image (see below).
Right:  Contours of the VLA 5\,GHz maps. Here only the active lobes A,B
are prominent, but we resolve the arcsec scale outflow. The background 
gray-scale shows the residual WIYN OPTIC $i^\prime$ image after removal 
of the host elliptical isophotes, shown with a linear intensity scale (see below).
}
\end{center}
\end{figure*}

\section{Observations}

\subsection{VLA Radio Imaging}

The mas- to arcsec-scale radio structure of 4C+37.11 has been
described by \citet{man04,rod06,rod09}.  The two compact cores C1 and
C2 suggested as the black hole nuclei lie at position angle
118$^\circ$ (measured N through E). The fainter, western nucleus C2 is
the origin of the CSO outflow, visible as two relativistically
expanding jets.  The northern jet at PA $\approx 25^\circ$ seems to
curve toward PA\,$\sim 0^\circ$ at the arcsecond scale. The southern
jet lies at angle $205^\circ$ out to $\sim 50$\,mas, then bends toward
$225^\circ$ at $\sim 5$\, arcsec. In the 4.9\,GHz map of \citet{man04}
there also appears to be a weaker axis of emission with extensions at
$PA\approx 80^\circ$ and $240^\circ$.

To trace these structures to arcsecond scales, we obtained 
VLA observations in the A configuration on November 11, 2012 (Proposal Code VLA/12B-058).
Integrations were obtained in bands centered at 1.3\,GHz (effective
bandwidth 384\,MHz, rms noise 50\,$\mu$Jy/beam, resolution $1.39
\times 1.25^{\prime\prime}$ at PA=$-81^\circ$) and 4.7\,GHz (effective
bandwidth 256\,MHz, rms noise 70\,$\mu$Jy/beam, resolution
$0.5^{\prime\prime}$).

All processing was performed in AIPS \citep{Greisen2003}.  Following editing
using RFLAG and after discarding some channels with low signal levels, we
fringe-fit the strong calibrator 3C286 to remove any residual delay
errors.  Next the data were amplitude calibrated using the AIPS tasks
SETJY, CALIB, and GETJY.  Bandpass calibration was determined from
3C286 while phase and polarization leakage calibration were obtained
from observations of the nearby calibrator J0414+3418.  Following 
calibration, imaging was carried out in AIPS taking advantage of 
new multi-frequency synthesis routines.

We show contours from the maps obtained in figure \ref{Radio}, overlayed on
the host emission at higher energy. The 4.7\,GHz contours show the
main CSO outflow from nucleus C2 (A, B) which appears to follow
optical `channels' in the host galaxy. At lower frequency the outflows
in these channels are unresolved, but the lobes (A, B) appear to
extend to $\sim 15^{\prime\prime}$ and occupy cavities in the host
X-ray map.  A second outflow axis (C, D) at roughly orthogonal
position angle also appears to correlate with X-ray cavities. This may
be an extension of the VLBA-measured secondary axis of \citet{man04}.

From a spectral index map made between 1.3 and 4.7 GHz (not shown), we
see that the core emission is very flat $\alpha \approx 0$, while the
lobes are much steeper $\alpha \approx$ $-$1.3 to $-$1.7, as seen in other
radio galaxies in dense environments.  The C,D structures are
particularly steep, being visible only at low frequency. The core of
4C+37.11 itself appears to be unpolarized with a limit derived from
Stokes Q and U images of $<0.2\%$. For the weaker lobes we constrain
the polarization to be $<1\%$. It is likely that this represents
de-polarization by the extensive cluster gas. We did observe
significant polarization in three background sources, but the nearest
of these was $\sim 8^\prime$ away, so this reveals little about the
cluster gas or magnetic field.

\subsection{WIYN Optical Imaging}

\begin{figure*}[ht!!]
\vskip 6.8truecm
\includegraphics{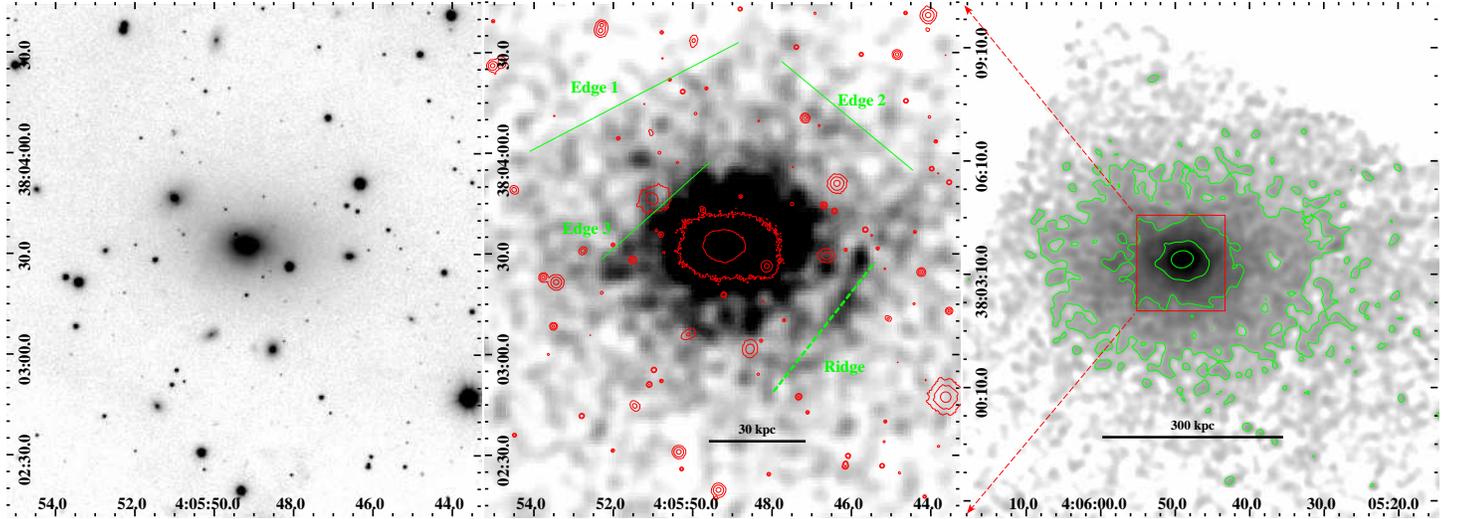}
\begin{center}
\caption{\label{OptX} 
Left: WIYN/OPTIC i' image of 4C+37.11.
Middle: 0.3-8\,keV ACIS-I image, smoothed with a 4$^{\prime\prime}$ Gaussian. The contours mark the
optical objects from the left frame, while lines indicate some diffuse X-ray features.
Right: Large scale 0.3-8keV emission, exposure corrected and smoothed with a 12$^{\prime\prime}$ Gaussian
kernel, log stretch. The position of the left optical/X-ray frame is shown by the red box. North up, East left.
}
\end{center}
\end{figure*}

	In Figure \ref{OptX} (left panel) we show a 2220\,s stack of $i^\prime$ exposures taken 
on Dec. 12 2008 using the OPTIC orthogonal frame-transfer camera \citep{tonry02,tonry04} 
on the Kitt Peak National Observatory, 3.5\,mm Wisconsin, Yale, Indiana, \& NOAO 
(hence WIYN) telescope. The frame-transfer allows rapid electronic guiding. The data
were reduced with standard IRAF routines \citep{tod86,val86}, except that the flat
field was dithered following the fast guiding corrections.
The dithering code, conflat2, was provided by J. Tonry (2011, private communication).
The combined image has a stellar FWHM resolution of 0.39$^{\prime\prime}$.
The host galaxy for 4C+37.11, centered in the frame, has an integrated $i^\prime$ magnitude
of 15.1. The next brightest galaxy lies 25.3$^{\prime\prime}$ (projected distance
26.7\,kpc) from the host core and is 2.5 magnitudes fainter at $i^\prime=17.6$. 
The next closest bright galaxies are at projected separations (and magnitudes) of 
28.2$^{\prime\prime}$ ($i^\prime=19.7$), 31.5$^{\prime\prime}$ ($i^\prime=18.6$) and
31.7$^{\prime\prime}$ ($i^\prime=18.5$).  While we do not have spectroscopic 
confirmation that these galaxies are associated, the optical and X-ray morphology do
show some evidence for interaction. For comparison, the middle panel shows diffuse 
X-ray emission at the same scale while the right panel of Figure 2 shows the 
large scale X-ray halo.

The \citet{schl98} maps give a large extinction $A_V=3.52$, $A_{i^{\prime}} =2.25$
in this direction. This gives a corrected host magnitude $i^\prime = 12.8$. The
2MASS data give an extinction-corrected total host magnitudes of $J=11.34$ and $K_s=9.94$.
Thus this is a very luminous galaxy, some $\sim 25\times$ more luminous
than $L^*$: the extinction-corrected absolute magnitudes are
$M_{i^\prime} = -24.1$ ($M^*_{i^\prime}-3.7$),
$M_{J} = -25.6$ ($M^*_J -3.3$),
$M_{K_s} = -27.0$ ($M^*_{K_s}-3.5)$.

\begin{figure}[t!!]
\vskip 8.2truecm
\includegraphics{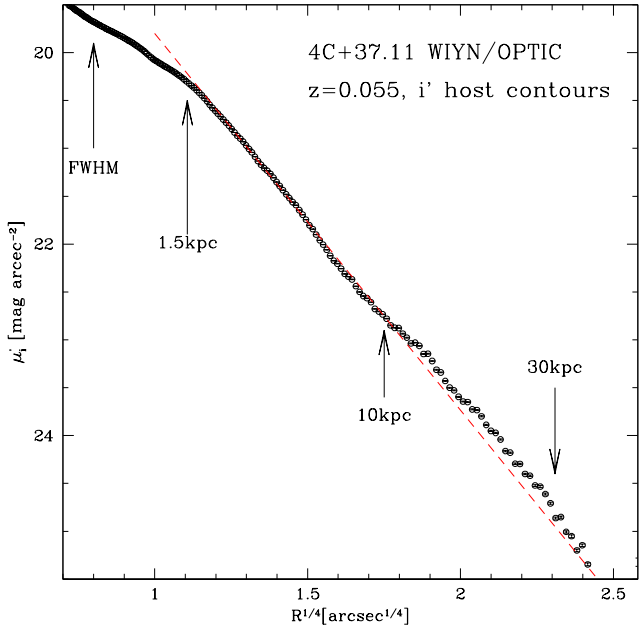}
\begin{center}
\caption{\label{deV} 
Host surface photometry averaged over elliptical isophotes. A reasonably good
approximation to a deVaucoleurs profile obtains to large radius, with no evidence
for a cD tail. The step at $\sim10$\,kpc shows the extent of the cavities
associated with the radio lobes. The inner kpc region shows strong flattening,
partly associated with the channels evacuated by AGN outflow.
}
\end{center}
\end{figure}

	We have performed an isophotal analysis of the host using the IRAF task `ellipse'.
The surface brightness (Figure \ref{deV}) shows a reasonable fit to a deVaucoleurs profile,
however there is a modest step in surface brightness at a radius of $\sim 10$\,kpc. Our
good resolution allows us to extend the analysis well within 1\,kpc, 
where we see an appreciable flattening of the galaxy core. Since there is no evidence
for a large excess to the $R^{1/4}$ law at large radii, we infer that this is not a cD galaxy.

	We have subtracted a smooth galaxy model generated from these elliptical
isophotes. Some elliptical residuals remain at $\sim 5^{\prime\prime}$ radii, but
the subtraction shows considerable real structure (Figure \ref{Radio}). Most prominent are low
surface brightness (here light) channels extending north and south to $\sim 10^{\prime\prime}$. 
These appear connected to and extend the VLA-scale radio lobes. A third channel 
extends to the E. All are tortuous. On smaller $<2^{\prime\prime}$ scales
there appear to be two white, but edge-brightened channels oriented
approximately E-W. The nature of these structures is unclear. They are 
approximately orthogonal to the VLBI-scale jet axis and so may represent 
equatorial obscuration. Alternatively, these may be channels evacuated by the
secondary flow axis visible in the VLA maps. Additional high resolution images,
with color information on sub-arcsec scales could help elucidate the nature
of these channels, distinguishing emission line structure and  dust obscuration from host cavities.

\subsection{Keck LRIS Spectroscopy} 

\begin{figure*}[ht!!]
\vskip 6.8truecm
\includegraphics{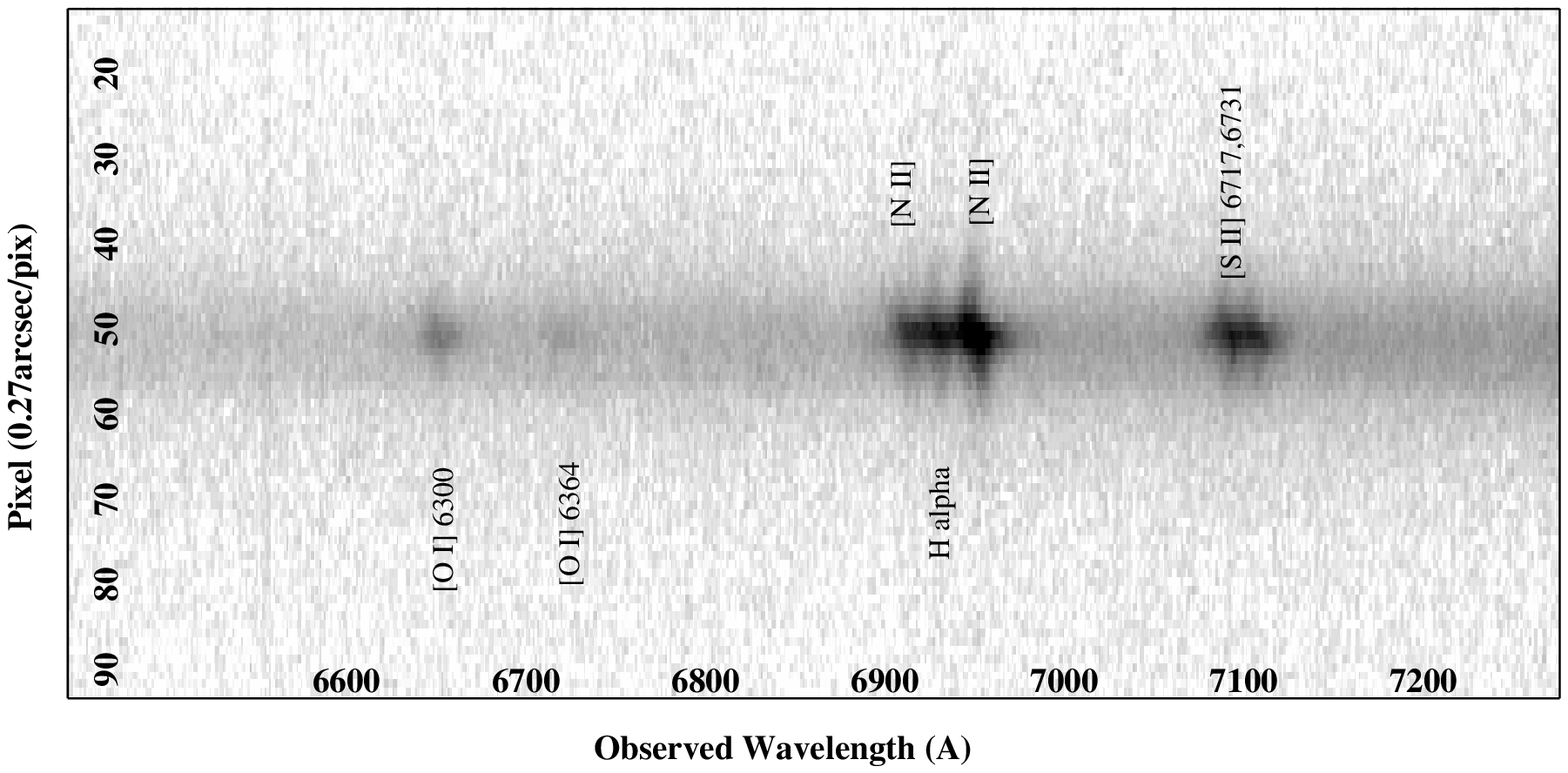}
\includegraphics{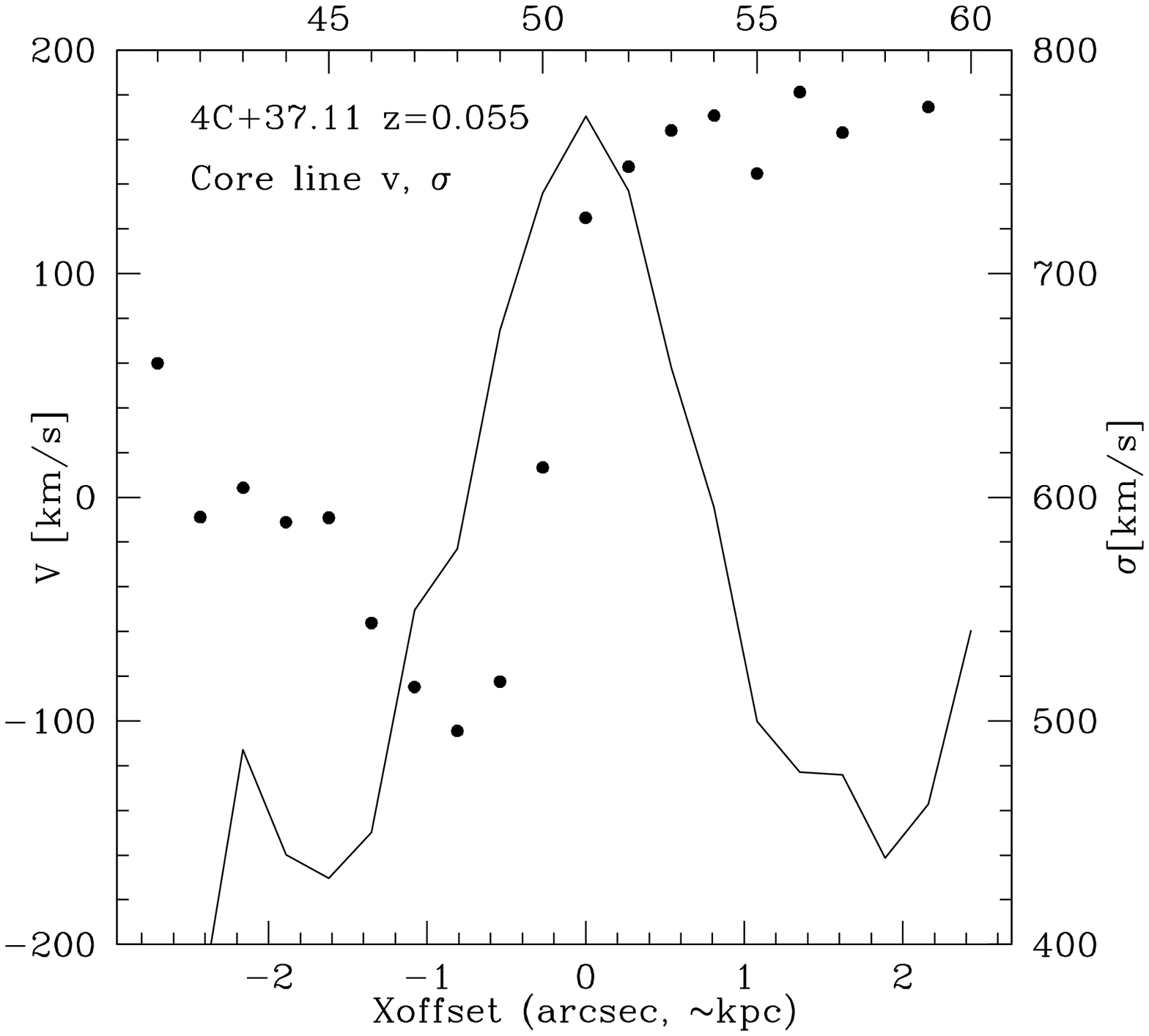}
\begin{center}
\caption{\label{2Dspec} 
Keck LRIS long slit spectra of the central region of 4C+37.11. The core shows
highly reddened forbidden line emission, which displays velocity shifts within
an arcsecond of the nucleus. The right panel shows the line shifts (dots, left scale) and 
broadening (line, right scale) of the H$\alpha$/N[II] complex. 
}
\end{center}
\end{figure*}

	Two exposures, totaling 1200\,s were obtained with the Keck I/LRIS on
Dec. 17, 2010. The red channel used the 400 line/8500\AA\, blaze grating for a 
dispersion of 1.18\AA/pixel. We used a 1 arcsec slit positioned approximately
along the VLBI jet axis (at PA=$20^\circ$, see right panel of Figure \ref{Radio}), giving 
an effective spectral resolution of 5.4\AA\, 
and a spatial resolution of 0.27$^{\prime\prime}$/(binned) pixel. With the 5600\AA\,
dichroic the red channel data covered 5500-10300\AA. The H$\alpha$/forbidden lines
near rest 6500\AA\, were well measured (Figure \ref{2Dspec}), and we detected the nebular 
[S III] 9069/9532\AA\, lines at 9570/10054\AA.
With the large extinction in this direction the spectra were highly reddened and 
little light was collected from the blue channel.

	The lines are resolved, with an S-shaped velocity profile along the
slit. To the north, the radial velocities are $\sim 300$km/s smaller than to the
south of the nucleus. The nature of these line shifts are not clear; to
the north the velocities seem to return to near the systemic velocity by
about 2$^{\prime\prime}$; to the south the offset persists. Since our position
angle is close to the host minor axis, the velocities probably do not represent
global rotation. They may instead represent velocity perturbations in the 
host interstellar medium
associated with the outflow along the VLBI-scale jet axis. This is at least consistent with
the VLBI jet flux ratio near the core \citep{rod09}, where the northern jet is brighter
near its base and thus may be emerging from the plane of the sky. In \citet{rod06} the
line splitting of $\sim 300$\,km/s was noticed in the $2^{\prime\prime}$ HET spectra,
and it was suggested that these might represent the radial velocities of the 
two BHB cores. At the projected separation of 7.3\,pc this indicated
an enclosed mass of $>1.5\times 10^8M_\odot$. It now seems likely that 
this integrated profile shape was caused by the larger scale outflows
resolved here. However, \citet{rod09}
found HI absorption velocities differing by $1000\,{\rm km\,s^{-1}}$ over
$\sim 7$\,mas in front of the southern jet, and used this to derive a lower mass
limit of $7 \times 10^8 M_\odot$.

	The core itself is not spatially resolved with the 
$\sim 1^{\prime\prime}$ seeing during our spectroscopic exposures,
but at the core the central velocity dispersion peaks at $\sim 750$\,km/s, after correcting
for the instrumental resolution. This suggests a mass of $\sim 7 \times 10^{10}
M_\odot$ in the central 0.5\,kpc, which easily accommodates the VLBI-scale mass
estimates above.
	
\subsection{{\it CXO} ACIS Imaging}

	As discussed in \citet{rod06}, 4C+37.11 was detected in the RASS and had two
HRI observations, showing an extended X-ray halo. More recently a short 3\,ks SWIFT 
XRT exposure confirmed diffuse
X-rays extending several arcmin from the host galaxy and allowed an estimate of 
$kT \approx 4$\,keV for the X-ray halo temperature. To improve the spectral measurements
and to get a first look at the arcsec-scale X-ray structure of 4C+37.11 we obtained an
exploratory {\it CXO} ACIS-I image on April 4, 2011 (ObsID 12704; Murray, PI).
Using a SIM offset the focal point and target were shifted onto the I3 chip, $>3^\prime$
from the array gaps so that the bulk of the diffuse halo would be covered on a single chip.
The TE exposures used the standard 3.1\,s frame time and was read out in VFAINT mode; we
were able to use the full 10\,ks exposure as there were no strong flaring events
during this observation.
This exposure netted $\sim 16,000$ counts (0.3-10\,keV) from the X-ray halo and $\sim 30$
counts from the AGN itself. 

	The data were subject to standard CIAO 4.3 processing. In Figure \ref{OptX}, right, we show
the large-scale distribution of the X-ray halo with a smoothed, exposure corrected
0.3-8\,keV image. The X-ray halo is elliptical with an axis ratio $a/b\sim 1.33$, and PA similar to
that of the central galaxy. It can be detected to a semi-major axis of at least $10^\prime$
(about 1100\,kpc, or 0.6$\times$ $r_{500}$, where the density drops to 500$\rho_{crit}$ as 
estimated in \S3), although we lack the S/N for a study beyond $\sim 7^\prime$. 
The inner X-ray isophotes are boxy, suggesting edges in the surface brightness. 
In the middle panel of Figure \ref{OptX}, we show the central 2.5 arcmin, stretched to emphasize some 
of the diffuse features. `Edge 1' passes 50$^{\prime\prime}$ NE of the central galaxy and
has a count rate surface brightness contrast of $\sim 2.5\times$. `Edge 2', 55$^{\prime\prime}$
to the NW is somewhat less well defined, but still seems to represent a $\sim 2\times$
step. `Edge 3', 20$^{\prime\prime}$ from the host core is centered just inside the 
nearest bright companion galaxy and with a $\sim 3\times$ intensity step may represent
a front associated with companion interaction. On the opposite SW side at 
$\sim 35^{\prime\prime}$ lies an apparent ridge of X-ray emission. Similar,
albeit less well defined, structure are visible at both larger (Figure \ref{OptX}, right)
and smaller (Figure \ref{Radio}, left) scale.  Unfortunately we lack sufficient counts 
to measure these structures with high accuracy or to determine if these are
shock fronts or cold fronts. However, clearly the halo of 4C+37.11 shows
extensive disturbance and abundant evidence of past interactions. A deeper X-ray
image is needed for an adequate spatial/spectral study of these structures.

	Turning to smaller scales, we can see a good correspondence between the 
radio outflow and the X-ray surface brightness (Figure \ref{Radio}). In particular the
northern radio lobe occupies a well-defined cavity in the X-rays. There also seems
to be decreased X-ray surface brightness along the E-W cavities of the secondary
outflow axis. However the southern radio lobe, which shows a marked cavity in the
optical, does not appear well defined in the X-ray.

\subsection{CXO X-ray Spectroscopy}

	For the AGN itself, we detect only 30$\pm$6 counts, embedded in a bright
structured background. These counts are hard with 1/3 of the background-subtracted photons
at $E>4$\,keV.  We can not distinguish heavy core absorption from hard spectrum nuclear emission
but if we assume
$N_H = 8 \times 10^{21} {\rm cm^{-2}}$, as determined from the halo emission (below), and 
a $\Gamma=1$ power law index a SHERPA fit gives a 2-10\,keV flux 
$6.3 \pm 2.4 \times 10^{-14} {\rm erg\,cm^{-2}s^{-1}}$. Thus we can only conclude that the emission
is consistent with that of a typical Seyfert core. In contrast for the diffuse halo
we have sufficient counts to estimate the temperature in a handful
of regions and to get the radial gradient in the emissivity.

	For an initial study of the halo, we defined a series of elliptical
shells centered on the AGN core, with semi-major axes 
$\theta_{maj} = (20\times 2^N)^{\prime\prime}$ and
semi-minor axes $0.75 \times \theta_{maj}$, with N taking values 0,1...4.
We fit various extraction apertures using {\it Sherpa}.
For all fits we used a background region $\sim 14^\prime$ from
the X-ray core on the I0 chip. This is along the halo minor axis at
a projected distance corresponding to $\sim 1.0-1.3 r_{500}$; while
this represents the maximum distance accessible with the active chips
there may be some suppression of the lowest surface brightness flux estimates.
Spectra were grouped to maintain S/N=3/bin and 
the source was assigned a model redshift z=0.055. All quoted errors are
1$\sigma$ single parameter values.

We first fit the sum of regions 2-4 (excluding the cooled core) to an 
absorbed APEC model, obtaining an overall temperature 
$kT_{sp} =4.8\pm0.2$\,keV and an
abundance (relative to Solar) of $Ab=0.63\pm 0.10$. The absorption was
found to be $N_H = 8.2 \pm 0.4 \times 10^{21} {\rm cm^{-2}}$. This
gives an equivalent $A_V= N_H/2.2\times 10^{21}{\rm cm^{-2}}=3.7$,
in excellent agreement with the full Galactic extinction estimated from the
Schlegel et al. (1998) maps, suggesting that the intrinsic absorption is small.
The fit source flux (out to $\theta_{maj}=320^{\prime\prime}$
($\approx 340$\,kpc $\approx r_{500}/3$) is 
$f_x ({\rm 0.5-8\,keV}) = 1.83\pm 0.07 \times 10^{-11} {\rm erg\,cm^{-2}s^{-1}}$.
The fit is quite acceptable with $\chi^2_{Geh}/$DoF=222/341.

\begin{figure}[hb!!]
\vskip 8.5truecm
\includegraphics{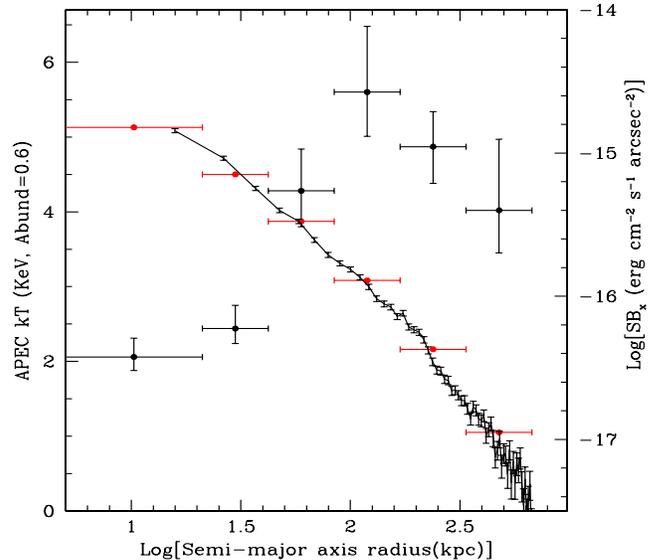}
\begin{center}
\caption{\label{radprof} 
The line shows the elliptical (aspect ratio 4:3) surface brightness profiles of the halo X-rays.
We also show the best-fit mean surface brightness (points with horizontal flags,
right scale) and temperature (horizontal and vertical flags, left scale) in our six
zones (Table 1). Note the strong cooling in the halo core.
}
\end{center}
\end{figure}

	We next fixed the extinction and abundance at the global values
and fit the temperature and flux in the elliptical zones using the
Sherpa {\it deproject} script, starting from the outermost shell. This
script accounts for the emission from larger shells to reconstruct
a true three-dimensional temperature and flux estimate.  We see that
the halo temperature peaks at about $2^\prime$ ($\sim 100$\,kpc Figure \ref{radprof}). We also show
the surface brightness determined by the de-projection fit to the 
elliptical shells and compare with a simple profile of the count 
rate/arcsec$^2$, where the latter has been normalized by the effective efficiency
6.9\,keV/cnt. While the halo is detectable to well beyond the
temperature peak, there are no dramatic steps in the azimuthal-averaged
surface brightness profile.

Although we lack the statistics to do a full de-projection analysis,
if we use simple annuli to fit the abundance we do detect a marginally
significant gradient with $Ab=1.0\pm0.3$ in the central cooling core,
and a drop off to $Ab=0.3\pm 0.2$ in the outermost regions. 

\begin{deluxetable}{lrllc}[t!!]
\tablecaption{\label{Params} X-ray Spectral Fits to the Halo of 4C+37.11}
\tablehead{
\colhead{N}&\colhead{$\theta_{maj}$(")}&\colhead{$kT$}(keV)&\colhead{$f_{0.5-8}$\tablenotemark{a}}&$\chi^2$/DoF 
}
\startdata
  0& 20& $2.05_{-0.18}^{+0.25}$ & $1.51_{-0.09}^{+0.09}$ & 69/112 \cr 
  1& 40& $2.44_{-0.20}^{+0.31}$ & $2.06_{-0.14}^{+0.12}$ & 68/127 \cr 
  2& 80& $4.28_{-0.48}^{+0.56}$ & $3.89_{-0.19}^{+0.22}$ & 98/195 \cr 
  3&160& $5.60_{-0.59}^{+0.88}$ & $6.04_{-0.25}^{+0.24}$ & 173/263 \cr 
  4&320& $4.87_{-0.49}^{+0.47}$ & $7.90_{-0.26}^{+0.26}$ & 146/324 \cr 
  5&640& $4.02_{-0.57}^{+0.95}$ & $4.30_{-0.28}^{+0.25}$ & 160/344 \cr 
\enddata
\tablenotetext{a}{Absorbed flux in $10^{-12} {\rm erg\,cm^{-2}s^{-1}}$}
\end{deluxetable}

	We have also attempted to fit the apparent edges shown in figure 1,
by estimating the temperature and flux in rectangular regions bracketing
the steps, but again lack the counts for a definitive characterization. For 
Edge 1 the spectra fits show (moving outward) a decrease in surface brightness
from $SB_X=3.9\pm0.3$ to $1.4\pm0.1$ (in units of $10^{-16} {\rm erg/cm^2/s/arcsec^2}$)
or a $2.8\pm0.3\times$ drop. The temperature step (from $5.8_{-1.2}^{+2.1}$\,keV to $6.4_{-1.3}^{+2.1}$)
was not however significant. In the core for Edge 3 we find a surface brightness
drop of $SB_X = 11.8\pm1.1 \longrightarrow 4.4\pm0.3$ ($2.7\pm0.3\times$) while 
the temperature increase $3.4_{-0.5}^{+0.6}$\,keV to $4.5_{-0.7}^{+1.0}$\,keV 
is only marginally significant.

\section{The Host Environment and Masses}

	It is useful to constrain the mass of 4C+37.11's host environment. The host
galaxy itself has $M_K=-27.0$ ($L=1.3 \times 10^{12} h_{70}^{-2} L_\odot$),
or $M\approx 9.5 \times 10^{11} M_\odot$. This is a massive elliptical. 
The dynamical time for this mass and a radius of $\sim$20 kpc is
about 60 Myr. There has been ample time for the stellar distribution
to be smoothed after a merger into the surface brightness profile seen
in our WIYN photometry.
The halo X-ray luminosity $1.2 \times 10^{44} {\rm erg\,s^{-1}}$ 
is also large. Thus the host components, while luminous, have a $L_X/L_{opt} \approx 40$ 
ratio not atypical for the brightest cluster galaxy \citep[BCG,][]{voe10}.

    The X-rays give a more direct measurement of the larger-scale potential.
\citet{vik06} have provided convenient scaling relations for low-$z$ clusters:
$$
r_{500} = 0.792 h^{-1} {\rm Mpc} (T_{sp}/5\,{\rm keV})^{0.53} \qquad {\rm and}
$$
$$
M_{500} = 2.89\times 10^{14} h^{-1} M_\odot (T_{sp}/5\,{\rm keV})^{1.58}.
$$
With our fit values and adopted cosmology these are $r_{500}= 1.1$\, Mpc and 
$M_{500}= 4\times 10^{14} M_\odot$.
Alternatively we can extrapolate from the extended flux within 400\,kpc
($\approx r_{2500}$, see Ascasibar et al 2006) to obtain an unabsorbed
$f_x (0.5-2{\rm keV}) = 2.1 \times 10^{-11} {\rm erg\,cm^{-2}s^{-1}}$
and an X-ray  luminosity $L_x (0.5-2{\rm keV}) = 1.5 \times 10^{44} {\rm erg\,s^{-1}}$.
For this $L_x$ the scaling relations in \citet{vik09} give 
$M_{500} \approx 1.4 \times 10^{14} M_\odot$. 

	\citet{lm04} have noted a correlation between the BCG K-band luminosity
and cluster mass $M_{200} \approx  1.36 M_{500}$
$$
M_{200} = (L_{BCG}/2.2 \times 10^{11} h^{-2} L_\odot)^{3.0\pm0.6}
10^{14} M_\odot .
$$
For our $K_s$ luminosity above this gives an exceptionally large 
$M_{500} = 1.9 \times 10^{15} M_\odot$. Since this is much larger than the direct
halo mass estimates, it suggests that the central galaxy may be somewhat anomalous
compared to the set studied in \citet{lm04}. Indeed the galaxy appears about 2.5$\times$
brighter than the trend-line in that study, so further measurements are needed to see if 
the host elliptical has, e.g. excess luminosity from ongoing star formation.

	The next question is the nature of this halo. Is it a typical galaxy cluster
or a `fossil' cluster? \citet{voe10} give the following criteria for a fossil:
for halos with $L_x>2.5\times 10^{41}h^{-2} {\rm erg\,s^{-1}}$, there should be
a `magnitude gap' $\Delta m_{12} > 1.7$ (in R) for all members within half 
the virial radius which we take as $0.7\,r_{500} \approx 700 {\rm kpc} \approx 12^{\prime}$. 
Certainly the nearby galaxies in our $i^\prime$ image satisfy this criterion. Given the
substantial extinction in this direction, a list of 2MASS extended
sources with total K-band magnitude $K_{tot} < 12.25$ within $0.7\,r_{500}$
is a useful catalog check. There are two such sources, 2MASS 04050165+3801503
with $K_{tot}= 11.8$ at 9.5$^\prime$ and 2MASS 04051504+3806532 with
$K_{tot}= 12.0$ at 7.5$^\prime$ projected separation. Neither has a redshift so
a spectroscopic exclusion of membership of (at least) these galaxies would be needed 
to confirm a fossil status for 4C+37.11.

	If it is a fossil, how long has it been since it's last major merger?
The presence of X-ray structure shows that this system is not fully relaxed.
However the central cavities, at least, can clearly be attributed to recent
activity by the central AGN. \citet{rod06} found velocities of $\sim 0.1$c
for the 20\,mas-scale radio lobes, indicating an age of $\sim 10^3$\,y. If
the bulk relativistic motion extends up to the full $\sim 5^{\prime\prime}$ scale
of the VLA radio lobes that would indicate a lifetime of $\sim 3 \times 10^5$y.
However, given that the apparent jet bending indicates slowing by $> 4 \times$
and that the $i^\prime$ host `channels' have a length approaching 
$30^{\prime\prime}$ it is likely that the source has remained
active for $>3\times 10^6$y.

	On larger scales, the X-ray halo shows evidence for edge structures. 
Unfortunately with the present S/N we cannot map their full extent, but such
structures can represent the `sloshing' of cluster gas, which, through comparison
with cluster hydrodynamics simulations, can be used to constrain the age of major merger
events \citep[and refs. therein]{jon11}.  Here the colder core gas of the main 
cluster, attracted by the sub-cluster's mass, takes a lagging orbit about the
overall potential minimum, making a sub-sonic trail, ram pressure confined
by the hotter halo gas from larger radii. A common sloshing pattern, for
finite impact parameter mergers close to the plane of the sky is a spiral
of such `cold' front edges. Such structures first form $\sim
0.3$\,Gyr after the first pass of the merging component
and can persist for several Gyr. On intermediate $\tau_{dyn}$ timescales, 
if a merger event is supersonic there can also be shocks showing interaction between
the main halo gas and that of a merging companion.

	For 4C+37.11, the general impression is a series of fronts 
(Edge1, Edge 3, the Ridge) normal to an axis at PA$\approx 45^\circ$. This
may represent a series of shells, rather than a large scale spiral. Such
structure could be the residue of a zero angular momentum merger close to the
plane of the sky or of a more typical merger spiral viewed nearly edge-on
with the angular momentum axis near the plane of the sky. In any event our
limited X-ray spectroscopic data do support the cold-front interpretation
with higher temperatures following lower surface brightness. However,
a deeper exposure is needed to see if this trend is a significant discontinuity
beyond the general cooling gradient. Also a more sensitive mapping of structure
at other position angles (e.g. `Edge 2') is needed to determine the
overall morphology and allow a comparison with simulations.

	We should finally check what sort of black hole mass we expect for this giant
elliptical host. \citet{gul09} have provided updates to the $M_\bullet -\sigma$
and $M_\bullet -L$ relations: 
$$
{\rm log}(M_\bullet/M_\odot) = 8.23\pm0.08 + (3.96\pm0.42){\rm log}(\sigma/200 {\rm km\,s^{-1}}),
$$
$${\rm log}(M_\bullet/M_\odot) = 8.95\pm0.11 + (1.11\pm0.18){\rm log}(L_V/10^{11}L_{\odot,v}).$$
For giant ellipticals at $z\approx 0.05$ we expect colors $V-K_s = 3.10$ and $V-i=0.64$.
Together with our absolute magnitude estimates, these give $V=-23.9$ and $V=-23.6$.
Thus our host luminosities suggest a (total) hole mass of ${\rm log}M_\bullet= 9.4-9.5$.
This may be easily accommodated by the central velocities measured by our long slit 
spectroscopy.

\section{Conclusions}

	4C+37.11 remains a unique source and a possible Rosetta stone for the
understanding of major mergers and the eventual fate of the component's
super-massive black holes (SMBHs). We have found here that the host of this SMBH binary
is a remarkably luminous elliptical galaxy and that in turn this is embedded
in a cluster-scale X-ray halo.  A dearth of other bright galaxies suggests that this
may be a `fossil' cluster. However, there are central structures in the X-ray
surface brightness that imply that this system is not fully relaxed. In addition we 
have discovered remarkable `channels' in the host galaxy optical surface brightness
that correlate well with the observed large scale radio outflow of the central engine.
Clearly this is a case where `feedback' is active. Interestingly, these channels 
are quite twisted, covering a large range of position angle about the central engine.
It is possible that this indicates a secondary outflow axis from the SMBHB or
a precession or spin-flip variation caused by SMBH interaction. In any event
it indicates relatively isotropic and, hence effective, mechanical heating of the
central X-ray halo.

	In addition we detect X-ray structures on larger $\sim 10$\,kpc scales.
At present it remains unclear if these represent `sloshing' cold fronts or
more recent shocks, but these likely preserve evidence of a past major merger
event.

	The next steps are to use high resolution color information to see whether the
host and its central channel structures show evidence of young stellar populations
or ISM emission. We can also improve our understanding of the sub-arcsecond
scale structure with additional, spatially resolved spectroscopy. Finally, on larger
scales, we could certainly profit from a deeper X-ray image. While the present 
snapshot suffices to show that there is rich X-ray structure and that it correlates
with known radio and optical morphology, we lack the counts to map this in detail
and to make a spectroscopic identification of the state of the gas. Given the 
unusual nature of this luminous halo and the possibility that it will shed light
on the unmerged state of the SMBHB at the core, this is a compelling quest,
and new observations to enable a detailed morphology/spectral study are underway.

\acknowledgments

This work was supported in part by NASA grants NNX10AD11G and
NNX10AP65G (RWR), NSF grant AST1109911 (ACR) and Chandra Award 
number GO2-13149X (GBT)
issued by the Chandra X-ray Observatory Center, which is operated by
the Smithsonian Astrophysical Observatory for and on behalf of NASA
under contract NAS8-03060.  We thank Adam van Etten for help with
the CIAO reductions.  R.T.Z. thanks M. Stickel for providing his
imaging data of 4C\,+37.11, S. Howell for assistance during the WIYN
runs and K. Herrmann for helpful discussions on narrow-band imaging.

Some of the data presented herein were obtained at the W.M. Keck Observatory, which 
is operated as a scientific partnership among the California Institute of Technology, the 
University of California and the National Aeronautics and Space Administration. The 
Observatory was made possible by the generous financial support of the W.M. Keck Foundation. 
The National Radio Astronomy Observatory is a facility of the National Science Foundation 
operated under cooperative agreement by Associated Universities, Inc.

{\it Facilities:} \facility{JVLA}. \facility{CXO (ACIS)}, \facility{WIYN (OPTIC), \facility{Keck:I (LRIS)}}


\begin{thebibliography}{}

\bibitem[Begelman, Blandford \& Rees(1980)]{bbr}Begelman, M.~C., Blandford, R.~D. \& Rees, M.~J. 1980, Nature, 287, 307
\bibitem[Bell et al.(2006)]{bet06}Bell, E.~F., et al.\ 2006, \apj, 652, 270
\bibitem[Boroson \& Lauer(2009)]{BL09}Boroson, T.~A. \& Lauer, T.~R. 2009, Nature, 458, 53
\bibitem[Burke-Spolaor(2011)]{BS11}Burke-Spolaor, S. 2011, MNRAS, 410, 2113
\bibitem[Cattaneo et al.(2009)]{cet09}Cattaneo, A., et al.\ 2009, Nature, 460, 213
\bibitem[Detweiler(1979)]{det79}Detweiler, S.\ 1979, \apj, 234, 1100
\bibitem[Eracleous et al.(2012)]{era12}Eracleous, M. et al.\ 2012, \apjs, 201, 23
\bibitem[Greisen(2003)]{Greisen2003}Greisen, E. W.\ 2003, in Information
Handling in Astronomy - Historical Vistas, ed. A. Heck, Astrophysics
and Space Science Library Vol. 285 (Dordrecht: Kluwer), 109
\bibitem[Gueltekin et al.(2009)]{gul09}Gueltekin, K., et al. 2009, ApJ, 698, 198
\bibitem[Hughes(2003)]{hug03}Hughes, S.~A. 2003, Annals of Physics, 303, 142
\bibitem[Johnson et al.(2011)]{jon11}Johnson, R.E., et al. 2011, ApJ, submitted; arXiv 1106.3489
\bibitem[Komossa et al.(2003)]{kom03}Komoosa, S. et al. 2003, ApJ, 582, L15
\bibitem[Lin \& Mohr(2004)]{lm04}Lin, Y.-T. \& Mohr, J.J. 2004, ApJ, 617, 879.
\bibitem[Maness et al.(2004)]{man04}Maness, H. L. et al. 2004, ApJ, 602, 123
\bibitem[Milosavljevic \& Merritt(2003)]{mm03}Milosavljevic, M. \& Merritt, D. 2003, ApJ, 560, 860
\bibitem[Owen et al.(1985)]{owen85}Owen, F.~N. et al. 1985, ApJ, 294, L85
\bibitem[Rodriguez et al.(2006)]{rod06}Rodriguez, C. et al. 2006, ApJ, 646, 49
\bibitem[Rodriguez et al.(2009)]{rod09}Rodriguez, C. et al. 2009, ApJ, 697, 37
\bibitem[Schlegel, Finkbinder \& Davis (1998)]{schl98}Schlegel, D.J., Finkbinder, D.P \& Davis, M. 1998, ApJ, 500, 525
\bibitem[Springel, DiMatteo \& Hernquist(2005)]{spr05} Springel, V., DiMatteo, T., \& Hernquist, L. 2005, MNRAS, 361, 776
\bibitem[Stickel, Kuehr \& Fried(1993)]{skf93} Stickel, M., Kuehr, H., \& Fried, J.~W.\ 1993, \aaps, 97, 483
\bibitem[Tody(1986)]{tod86}Tody, D. 1986, SPIE, 627, 733
\bibitem[Tonry et al.(2002)]{tonry02} Tonry, J.~L., Lupino, G.~A.,
Kaiser, N., Burke, B. \& Jacoby, G.~H. 2002, Proc. SPIE, 4836, 206
\bibitem[Tonry, Burke \& Schechter(1997)]{ton97}Tonry, J., Burke, B.E. \& Schechter, P.L. 1997, PASP, 109, 1154
\bibitem[Tonry et al.(2004)]{tonry04} Tonry, J.~L., Burke,
B.~E., Luppino, G., \& Kaiser, N.\ 2004, in ASSL 300, Scientific Detectors for Astronomy,
The Beginning of a New Era, ed. P. Amico, J.~W. Beletic, J.~E. Beletic,
(New York, NY: Kluwer), 385
\bibitem[Valdes(1986)]{val86}Valdes, F. 1986, SPIE, 627, 749
\bibitem[Valdes(1992)]{val92}Valdes, F. 1992, ASPconf, 25, 417
\bibitem[Vikhlinin et al.(2006)]{vik06}Vikhlinin, A. et al. 2006, ApJ, 640, 691
\bibitem[Vikhlinin et al.(2009)]{vik09}Vikhlinin, A. et al. 2009, ApJ, 692, 1033
\bibitem[Voevodkin et al(2010)]{voe10}Voevodkin, A., et al 2010, ApJ, 708, 1376
\bibitem[Voges et al.(1999)]{rosat} Voges, W., Aschenbach, B., Boller, T., et al.\ 1999, \aap, 349, 389 
\end{thebibliography}
\end{document}